\documentclass[english]{article}
\usepackage[T1]{fontenc}
\usepackage[latin9]{inputenc}
\usepackage{geometry}
\usepackage{xcolor}
\usepackage{array}
\usepackage{multirow}
\usepackage{graphicx}
\usepackage{float}
\usepackage{authblk}
\usepackage{amsmath}
\usepackage{soul}
\usepackage{textcmds}
\usepackage{graphicx}
\usepackage[utopia]{mathdesign}
\makeatletter

\begin{document}

\title{Controlled-Joint Remote Implementation of Operators and its Possible Generalization}
\author[1,$\#$]{Satish Kumar}
\author[2,3,$\star$]{Nguyen Ba An}
\author[1,$\ast$]{Anirban Pathak}
\affil[1]{Jaypee Institute of Information Technology, A 10, Sector 62, Noida, UP-201309, India}
\affil[2]{Thang Long Institute of Mathematics and Applied Sciences, Thang Long University, Nghiem Xuan Yem, Hoang Mai, Hanoi, Vietnam}
\affil[3]{Center for Theoretical Physics, Institute of Physics, Vietnam Academy of Science and Technology, 18 Hoang Quoc Viet, Cau Giay, Hanoi, Vietnam}
\affil[$\ast$]{Corresponding author: anirban.pathak@gmail.com}
\affil[$\#$]{mr.satishseth@gmail.com}
\affil[$\star$]{annb@thanglong.edu.vn}

\maketitle

\begin{abstract}
    The existing notion of the shared entangled state-assisted remote preparation of unitary operator (equivalently the existing notion of quantum remote control) using local operation and classical communication is generalized to a scenario where under the control of a supervisor two users can jointly implement arbitrary unitaries (one unknown unitary operation by each or equivalently a single unitary decomposed into two unitaries of the same dimension and given to two users) on an unknown quantum state available with a geographically separated user. It is explicitly shown that the task can be performed using a four-qubit hyperentangled state, which is entangled simultaneously in both spatial and polarization degrees of freedom of photons. The proposed protocol which can be viewed as primitive for distributed photonic quantum computing is further generalized to the case that drops the restrictions on the number of controllers and the number of parties performing unitaries and allows both the numbers to be arbitrary.  It is also shown that all the existing variants of quantum remote control schemes can be obtained as special cases of the present scheme.\\
       
\end{abstract}

\section{Introduction}
Entanglement is known to be an important resource for quantum computing and communication. The importance of the entangled state underlies in the fact that it (along with a slightly stronger version of it called nonlocal states)  can be used to perform various tasks that cannot be done in the classical world. For example, restricting us to the context of the present work, we may mention that entanglement can be used as a resource to realize the so-called quantum teleportation- an idea introduced by Bennett et al. in 1993 \cite{BB93}. In a conventional teleportation scheme, an unknown single-qubit quantum state is transferred from one place to another far away place without physically sending the qubit itself with the help of a shared bipartite entangled state, local operations and 2 bits of classical communication. Later, Pati et al. \cite{P00} showed that a known quantum states can be teleported to a receiver using an entangled state and a classical bit only. Such a scheme for teleportation of a known quantum state is known as quantum remote state preparation (RSP). Subsequently, the introduction of the concept of RSP in 2000 led to two different kinds of research interests. On one hand, several variants of RSP (e.g., controlled remote state preparation, joint remote state preparation, controlled joint remote state preparation, bidirectional remote state preparation have been proposed (see \cite{SSB+15} and references therein), on the other hand, a dedicated effort has been made to address the question: In analogy to RSP can we remotely prepare a quantum operation? The question was answered in the affirmative by Huelga et al. in 2001 \cite{HVC+01} and it was shown that quantum operation can be prepared remotely using shared entanglement along with local operation and classical communication (LOCC). Such a remote realization of quantum operations using shared entanglement is referred to as the quantum remote control (some authors have referred to it as the remote implementation of an operator (RIO), too). Before we proceed further it would be apt to note that any scheme for bidirectional quantum state teleportation \cite{TVP15,SST+17} can be trivially used for implementing a scheme for RIO. This can be visualized easily if we consider that Bob wishes to implement an arbitrary operator $U_B$ remotely on a quantum state $|\psi\rangle$ available with Alice. Now, Alice may teleport the state $|\psi\rangle$ to Bob and he may apply his operation on the state received to yield $|\psi^{\prime}\rangle=U_B|\psi\rangle$ and teleport the state $|\psi^{\prime}\rangle$ to Alice. This trivial scheme would require at least two copies of Bell states and 4 bits of classical communication. This sets a kind of upper limit on the resource requirement as the intentional use of a higher amount of resource will make no sense. Now, a scheme of efficient RIO would require a lesser amount of resources, and in Ref. \cite{HPV02}, it's shown that RIO can be implemented using two copies of Bell states and 4 bits of classical communication, and the same is a minimal requirement. In \cite{HPV02}, it was also shown that if $U_B$ mentioned above belongs to certain classes of unitary operators then the task (i.e., RIO) can be implemented using 1 Bell state and 2 bits of classical communication only. In 2006 one of the present authors came up with an idea of remote implementation of a hidden operator \cite{A07}: the necessary operator is hidden in a lump operator given to the implementer who can locally manipulate the lump operator as a whole only. Several variants of such remote implementation of hidden operator have also been proposed such as controlled remote implementation of
partially unknown quantum operation \cite{FL08}, cyclic controlled remote implementation
of partially unknown quantum operations \cite{PH19} and double-direction cyclic controlled remote implementation of partially known quantum operations \cite{PY+22}. It is interesting to note that a specific version of RIO is experimentally realized. Specifically, remote implementation of a rotation angle was experimentally demonstrated in \cite{XLG05}.

Since the introduction of the concept of quantum remote control, several variants of it have been proposed that are analogous to the variants of RSP. For example, controlled remote implementation of operator (CRIO) \footnote{Note that if we use particle order permutation technique as described and utilized in \cite{TVP15} and the scheme of \cite{HPV02}, a scheme of CRIO of an arbitrary operator would require 2 Bell states and 4 bits of classical communication, whereas a trivial scheme for CRIO obtained by modifying an efficient scheme for controlled bidirectional quantum teleportation \cite{TVP15} would require one more classical bit.} \cite{BB22} and joint remote implementation of operator (JRIO) \cite{BAn22} have been proposed. The proposed schemes utilize different types of quantum resources. For example, in \cite{BB22,BAn22,WG24} hyperentanglement is used for RIO, CRIO and JRIO and in \cite{QC23} graph state is used for CRIO. Thus, the schemes for RIO, CRIO, and JRIO have already been studied with reasonable rigor, but no scheme for controlled joint remote implementation of operator (CJRIO) has yet been proposed. This has motivated us to look into the possibility of designing such a protocol. Also, there is an additional motivation, the scheme for CJRIO  can be easily reduced to the schemes for JRIO, RIO, and CRIO. Further, such a scheme can be of use in distributed quantum computing requiring nonlocal operation (see \cite{GP07} and references therein) as well as in the quantum networks in general and quantum internet in particular.  For example, in Ref. \cite{CCC20} a device architecture for distributed quantum computing is proposed which is very apt for the present situation where only noisy intermediate quantum computers are available. As the available quantum computers are small in size, RIO will be essential in all such situations where the number of qubits required to perform a specific computational task exceeds the number of qubits that can be stored and compiled in a single quantum computer \cite{CCC20}. Now, as the task can be distributed over a large number of small quantum computers, JRIO is a requirement. Further, a master-slave architecture is often used in traditional distributed computing, where a master node (user) acts as the central control unit that receives tasks from clients and distributes the task among slave nodes. In our situation, slave nodes are located at different quantum computers and the master node is referred to as the controller, leading to an analogous situation in the quantum world leading to the requirement of CJRIO. Now, in the classical world, remote operations mentioned here are usually referred to as teleoperations and there exist schemes for teleoperations that involve multiple masers (i.e., controllers in our case) \cite{MFP+19}. A quantum analog of such a scheme would be a generalized version of CJRIO allowing multiple controllers (master nodes). Interestingly, this need and the fact that no scheme for CJRIO (independent of the number of controllers) exists, motivated us to design a scheme for CJRIO with a single controller first and then to generalize that to a multiple controller situation.

In this paper, we have first proposed a scheme for CJRIO using an entangled state and LOCC.  It is explicitly shown that CJRIO can be realized using a four-qubit hyperentangled state, which is entangled in spatial and polarization degree of freedom of photons. The preparation of such a hyperentangled state can be found in references \cite{DR+17Hyperentangle,YC24Hyperentangle,NL+23Hyperentangle,ZY+23Hyperentangle}. The proposed protocol is also generalized to the case that allows an arbitrary number of controllers and an arbitrary number of parties to perform unitaries. This is not only the most generalized version of quantum remote control, it can also be reduced to all the existing variants of quantum remote control schemes. Specifically, schemes for RIO, CRIO, JRIO, etc. can be obtained as special cases of the CJRIO schemes proposed here.

The rest of the paper is organized as follows. In Section \ref{sec:the task}, we briefly describe the task that we wish to perform here. Thereafter, in Section \ref{sec:CJRIO protocol}, we propose a scheme for CJRIO using a four qubit hyperentangled state, which is entangled at the same time in double degrees of freedom -the spatial and the polarization ones. Subsequently, in Section \ref{sec:generalization} we have generalized our protocol to the case where any number of parties can jointly prepare the quantum unitary in the supervision of an arbitrary number of controllers. The process for reducing our proposed scheme for CJRIO into the existing variants of RIO is described in Section \ref{RIOasCJRIO}. Finally, in Section \ref{sec:conclusion} the relevance of the proposed protocols is discussed and the paper is concluded.

\section{The task of interest}\label{sec:the task}
The idea of CJRIO is to jointly and controllably operate an unknown quantum operation on an unknown quantum state at different nodes. Here, we consider that an arbitrary unitary $U$ which can be decomposed as $U=U^1_B.U^2_B$ is implemented by Bob$^1$ and Bob$^2$ jointly on an arbitrary quantum state $ |\psi\rangle_X$ available with Alice who is spatially separated from Bob$^1$ and Bob$^2$. For a generalized view, we consider that the operators which Bob$^1$ and Bob$^2$ wish to operate are $U^1_B$ and $U^2_B$ respectively, the form of which can be given as follows:
\begin{equation}\label{eq:Ub1}
    U^1_B=\begin{pmatrix}
    u^1_B & v^1_B\\ 
    -v^{*1}_B & u^{*1}_B
    \end{pmatrix}
\end{equation}
\begin{equation}\label{eq:Ub2}
    U^2_B=\begin{pmatrix}
    u^2_B & v^2_B\\ 
    -v^{*2}_B & u^{*2}_B
    \end{pmatrix}
\end{equation}

As mentioned above, Alice who is spatially separated from Bob$^1$ and Bob$^2$ has an unknown quantum state $ |\psi\rangle_X$ of the following form:
\begin{equation}\label{eq:Qs}
    |\psi\rangle_X=(\alpha|x_0\rangle+\beta|x_1\rangle)_X|\text{V}\rangle_X
\end{equation}
where $\alpha$ and $\beta$ are unknown coefficients which satisfies the normalization condition $|\alpha|^2+|\beta|^2=1$, and $ |V\rangle_X$ describing the polarization state of photon. Physically, it can be viewed as if Alice has photon indexed by $X$ which is vertically polarized and is in spatial superposition state of $|x_0\rangle$ and $|x_1\rangle$.\\

The action of unitary operator $U^1_B$ on $|\psi\rangle_X$ can be described  as $|\psi_{B^1}\rangle=U^{1}_B|\psi\rangle_X=\alpha_{B^1}|x_0\rangle+\beta_{B^1}|x_1\rangle$ with $\alpha_{B^1}=\alpha{u^1_B}+\beta{v^1_B}$ and $\beta_{B^1}=-\alpha {v^{*1}_B}+\beta{u^{*1}_B}$. Further, the action of unitary operator $U^2_B$ on $|\psi\rangle_X$ can be described as $|\psi_{B^2}\rangle=U^{2}_B|\psi\rangle_X=\alpha_{B^2}|x_0\rangle+\beta_{B^2}|x_1\rangle$ with $\alpha_{B^2}=\alpha{u^2_B}+\beta{v^2_B}$ and $\beta_{B^2}=-\alpha {v^{*2}_B}+\beta{u^{*2}_B}$.\\

The task of concern is that Bob$^1$ and Bob$^2$ should remotely apply their operators on Alice's state, which can be mathematically represented as
\begin{equation}\label{eq:task}
\begin{split}
    |\psi_{B^1B^2}\rangle & ={U^1_B}{U^2_B}|\psi\rangle_X\\
    & = (\alpha_{B^1 B^2}|x_0\rangle+\beta_{B^1 B^2}|x_1\rangle)|\text{V}\rangle
\end{split}
\end{equation}
where, $\alpha_{B^1B^2}=\alpha_{B^2}{u^1_B}+\beta_{B^2}{v^1_B}$ and $\beta_{B^1B^2}=-\alpha_{B^2}{v^{*1}_B}+\beta_{B^2}{u^{*1}_B}$.  Please check this\\

The quantum channel used here to perform the task of concern is a four-qubit hyperentangled state given as
\begin{equation}\label{eq:QC}
    |Q\rangle_{AB^1B^2C}=|Q^S\rangle_{AB^1B^2C} |Q^P\rangle_{AB^1B^2C}
\end{equation}
where 
\begin{equation}\label{eq:6}
    |Q^S\rangle_{AB^1B^2C}=|a_0\rangle_A |b^1_0\rangle_{B^1}
|b^2_0\rangle_{B^2} |c_0\rangle_C + |a_1\rangle_A |b^1_1\rangle_{B^1}
|b^2_1\rangle_{B^2} |c_1\rangle_C
\end{equation}
\begin{equation}\label{eq:7}
    |Q^P\rangle_{AB^1B^2C}=|\text{H}\rangle_A |\text{H}\rangle_{B^1}
|\text{H}\rangle_{B^2} |\text{H}\rangle_C + |\text{V}\rangle_A |\text{V}\rangle_{B^1}
|\text{V}\rangle_{B^2} |\text{V}\rangle_C
\end{equation}
with $a_j$, $b^1_j$, $b^2_j$, $c_j$ $(j = 0,1)$ the spatial paths while $\text{H}$ and $\text{V}$ the horizontal and vertical polarization. The superscript $S$ denotes the spatial degree of freedom (S-DOF) and $P$ denotes the polarization degree of freedom (P-DOF). It is to be noted that in Eq. \ref{eq:6} and Eq. \ref{eq:7} the factor of normalization $1/\sqrt{2}$ is omitted. The labeling $A$, $B^1$, $B^2$ and $C$ in the subscript denotes photon state with Alice, Bob$^1$, Bob$^2$, and Charlie respectively. Bob$^1$ and Bob$^2$ try to implement an arbitrary unitary operation on an unknown state at Alice's node with all the participants physically far apart from each other.   

\section{Protocols for Controlled-Joint Remote Implementation of Operators\label{sec:CJRIO protocol}}

The combined state of Alice's state and quantum channel can be written as
\begin{equation}\label{eq:totalS}
    |\psi\rangle_X|Q^{SP}\rangle_{AB^1B^2C}=|\phi^S\rangle_{XAB^1B^2C}|\text{V}\rangle_{X}|Q^{P}\rangle_{AB^1B^2C}
\end{equation}
where
\begin{equation}\label{eq:9}
    |\phi^{S}\rangle=(\alpha|x_0\rangle+\beta|x_1\rangle)_X 
\otimes(|a_0\rangle_A |b^1_0\rangle_{B^1}
|b^2_0\rangle_{B^2} |c_0\rangle_C + |a_1\rangle_A |b^1_1\rangle_{B^1}
|b^2_1\rangle_{B^2} |c_1\rangle_C)    
\end{equation}
We will now only consider S-DOF of the combined state and will come back to P-DOF in Sec. \ref{P-DOF}.
\subsection{Utilizing S-DOF\label{S-DOF}}
\begin{figure}
    \centering
    \includegraphics[scale=1.2]{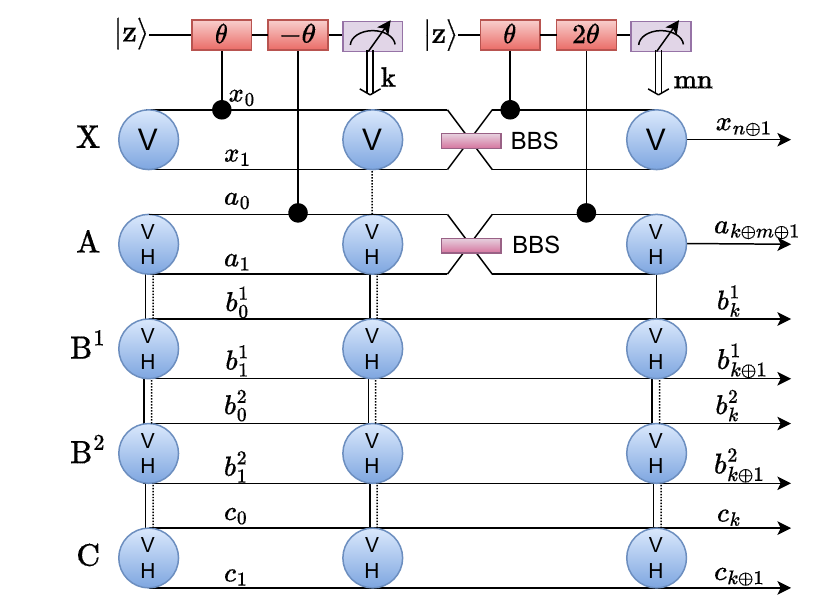}
    \caption{(Color online) A schematic which represents first two steps of the CJRIO protocol. A circle with V, H represents photon simultaneously in vertical and horizontal polarization and circle with V only represents photon in vertical polarization. The two (one) lines attached with circles represent photons having two spatial paths simultaneously (photons having one path only). The cross-kerr nonlinear interaction between a photon path and the coherent state is attached by a line with a bold dot on the photon path. The dimensionless parameter $\theta$ determines the change of phase of the CS brought about by the cross-kerr interaction. The double arrow from the coherent state represents measurement outcomes. Vertical solid (dashed) line represents entanglement in P-DOF (S-DOF). BBS here is a balanced beam splitter. Here, the photon $\text{X}$ first gets entangled with remaining photons by allowing non-linear interaction $K_{x_0}(\theta)|z\rangle|x_0\rangle.$ and $K_{a_0}(-\theta)|z\rangle|a_0\rangle.$ The measurement of the CS gives outcome $k$. Once photon $\text{X}$ gets entangled, then Alice tries to pass the coefficient of $|\psi\rangle$ to joint parties by mixing her photon paths on BBSs and allowing the interaction $K_{x_0}(\theta)|z\rangle|x_0\rangle$ and $K_{a_k}(2\theta)|z\rangle|a_0\rangle$ then measures the CSs, whose measurement outcomes are $m$ and $n$.}
    \label{fig:cjrio1}
\end{figure}

\begin{figure}
    \centering
    \includegraphics[width=\textwidth]{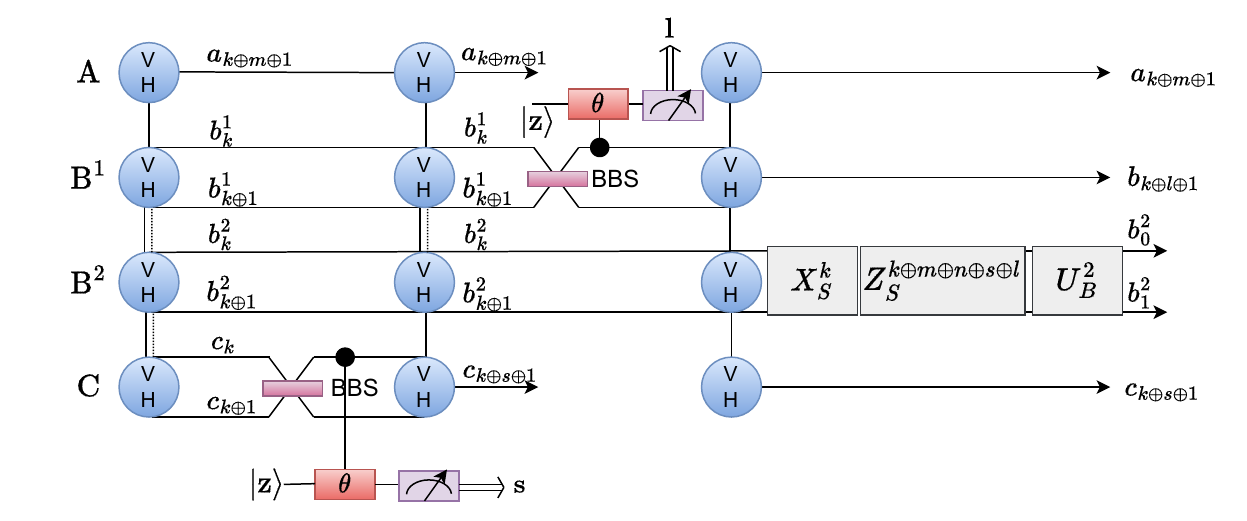}
    \caption{(Color online) A schematic which represents Step3 and Step4 of the CJRIO protocol. Here, controller Charlie first mixes her photon paths on a BBS and allows the interaction $K_{c_k}(\theta)|z\rangle|c_k\rangle$ and measures the CS, whose measurement outcome is $s$, which disentangles photon $\text{C}$ from remaining photons in S-DOF. After that Bob$^1$ mixes his photon paths on a BBS and allows the interaction $K_{b^1_k}(\theta)|z\rangle|b^1_k\rangle$ and measures the CS, whose measurement outcome is $l$, which allows Bob$^2$ to apply appropriate unitaries to get $\alpha_{B^2}|b^2_0\rangle+\beta_{B^2}|b^2_1\rangle$.}
    \label{fig:cjrio2}
\end{figure}

\begin{description}
    \item[Step1] The first step is to entangle photon $X$ with the remaining photons of the quantum channel. To do so, Alice prepares an auxiliary coherent state (CS) $|z\rangle$ and lets it interact with one of the path of photon $X$ (here $|x_0\rangle$) and photon $A$ (here $|a_0\rangle$) via cross-kerr nonlinear interaction\footnote{The cross-kerr nonlinear interaction between an auxilary coherent state $|z\rangle$ ($|z\rangle=\text{exp}(-|z|^2/2)\sum_{n=0}^{\infty}(z^n/\sqrt{n!}|n\rangle)$ where $|n\rangle$ is a Fock state containing $n$ photons) and a photon path, lets say $|b\rangle$ with interaction parameters $\theta$ and $-\theta$ is mathematically represented as $K_b(\pm \theta)|z\rangle|b\rangle=|ze^{\pm i\theta}\rangle|b\rangle$. The X-quadrature homodyne detection technique is used to measure whether the coherent state is in $|z\rangle$ or $|ze^{\pm i\theta}\rangle$. It is to be noted that $|ze^{i\theta}\rangle$ and $|ze^{-i\theta}\rangle$ are indistinguishable with this kind of measurement.} with interaction parameters $\theta$ and $-\theta$ ($K_{x_0}(\theta)$ and $K_{a_0}(-\theta)$) respectively. The measurement of the coherent state gives two possible outcomes $k=0$ $(1)$ corresponding to $|z\rangle$  $(|ze^{\pm i\theta}\rangle)$. Alice's measurement outcome changes the state in Eq. \ref{eq:9} to the following: 
    \begin{equation}\label{eq:10}
        |\xi_k\rangle=\alpha|x_0\rangle_X |a_k\rangle_A |b^1_k\rangle_{B^1}
|b^2_k\rangle_{B^2} |c_k\rangle_C + \beta|x_1\rangle_X |a_{k\oplus1}\rangle_A |b^1_{k\oplus1}\rangle_{B^1}
|b^2_{k\oplus1}\rangle_{B^2} |c_{k\oplus1}\rangle_C
    \end{equation}
with $\oplus$ denoting an addition mod $2$. Now, one can see from Eq. \ref{eq:10} that the photon $X$ is entangled with the remaining photons in the quantum channel.

    \item[Step2] Then, Alice tries to disentangle her photons $X$ and $A$ from the remaining photons in S-DOF. To do so, Alice mixes the two spatial paths $|x_0\rangle$ and $|x_1\rangle$ (|$a_k\rangle$ and $|a_{k\oplus1}\rangle$) of her photon $X$ ($A$) on a balanced beam splitter (BBS). The BBS transformation rule is (up to a normalization factor) $|\sigma_j\rangle\rightarrow|\sigma_j\rangle+(-1)^j|\sigma_{j\oplus1}\rangle$. After mixing the photons on the BBS, the state in Eq. \ref{eq:10} reduces to:
    \begin{equation}\label{eq:11}
    \begin{split}
|\xi^{'}_k\rangle & =(|x_0\rangle|a_k\rangle+(-1)^k|x_1\rangle|a_{k+1}\rangle)\otimes(\alpha|b^1_k\rangle||b^2_k\rangle|c_k\rangle+(-1)^k\beta|b^1_{k+1}\rangle|b^2_{k+1}\rangle|c_{k+1}\rangle)\\
&+(|x_0\rangle|a_{k+1}\rangle+(-1)^k|x_1\rangle|a_{k}\rangle)\otimes(\alpha|b^1_k\rangle||b^2_k\rangle|c_k\rangle-(-1)^k\beta|b^1_{k+1}\rangle|b^2_{k+1}\rangle|c_{k+1}\rangle)
    \end{split}
    \end{equation}
   It can be seen from Eq. \ref{eq:11} that photons $X$ and $A$ are still not separated from the remaining photons in the channel. To get them separated, Alice again uses an auxiliary CS $|z\rangle$ and turns on the cross-kerr nonlinear interaction with photon $X$ on path $|x_0\rangle$ and photon $A$ on path $|a_k\rangle$ with interaction parameters $\theta$ and $2\theta$ respectively. Alice then measures the X-quadrature of the CS whose measurement outcomes are $mn=00$, $01$, $10$ and $11$ corresponding to $|z\rangle$, $|ze^{i\theta}\rangle$, $|ze^{i2\theta}\rangle$ and $|ze^{i3\theta}\rangle$ respectively. After the measurement, the state becomes: 
\begin{equation}\label{eq:12}
    |\xi_{kmn}\rangle=|x_{n\oplus1}\rangle|a_{k\oplus m\oplus1}\rangle(\alpha|b^1_k\rangle|b^2_k\rangle|c_k
    \rangle+(-1)^{k\oplus m\oplus n}\beta|b^1_{k\oplus1}\rangle|b^2_{k\oplus1}\rangle|c_{k\oplus1}\rangle)
\end{equation}
The photons $X$ and $A$ are now separated in S-DOF from the remaining photons in the channel, which can be seen from Eq. \ref{eq:12}. To avoid complexity, we may forget photon $X$ and the Eq. \ref{eq:12} can be written as:
\begin{equation}\label{eq:13}
    |\Xi_{kmn}\rangle=|a_{k\oplus m\oplus1}\rangle(\alpha|b^1_k\rangle|b^2_k\rangle|c_k
    \rangle+(-1)^{k\oplus m\oplus n}\beta|b^1_{k\oplus1}\rangle|b^2_{k\oplus1}\rangle|c_{k\oplus1}\rangle)
\end{equation}
\end{description}
The Step1 and Step2 have been presented in pictorial form in Fig. \ref{fig:cjrio1}. Photons are labeled as $\text{X}$, $\text{A}$, $\text{B}^1$, $\text{B}^2$ and $\text{C}$. Photons $\text{X}$ and $\text{A}$ are with Alice and photons $\text{B}^1$, $\text{B}^2$ and $\text{C}$ are with Bob$^1$, Bob$^2$ and Charlie respectively. 
\begin{description}
\item[Step3] Now let us understand of the role of controller Charlie. If Charlie wants to stop the joint operation then she does nothing otherwise she mixes her photon path states $|c_k\rangle$ and $|c_{k\oplus1}\rangle$ on a BBS which transforms Eq. \ref{eq:13} into the following:
\begin{equation}\label{eq:14}
    \begin{split}
      |\Xi^{'}_{kmn}\rangle &= |a_{k\oplus m\oplus1}\rangle[(\alpha|b^1_k\rangle|b^2_k\rangle+(-1)^{m\oplus n}\beta|b^1_{k\oplus1}\rangle|b^2_{k\oplus1}\rangle)|c_k\rangle\\
      &+(-1)^k(\alpha|b^1_k\rangle|b^2_k\rangle-(-1)^{m\oplus n}\beta|b^1_{k\oplus1}\rangle|b^2_{k\oplus1}\rangle)|c_{k\oplus1}\rangle]  
    \end{split}
\end{equation}
She will then take an arbitrary CS $|z\rangle$ and let it interact with one of her photon path states $|c_k\rangle$ via cross-kerr nonlinear interaction with interaction parameter $\theta$ and then measure it. Let the measurement outcome be $s=0$ $(1)$ corresponding to $|z\rangle$ $(|ze^{i\theta}\rangle)$. After the measurement, the state in Eq. \ref{eq:14} transforms into the following:  
\begin{equation}\label{eq:15}
    |\Xi_{kmns}\rangle=|a_{k\oplus m\oplus1}\rangle(\alpha|b^1_{k}\rangle|b^2_k\rangle-(-1)^{m\oplus n\oplus s}\beta|b^1_{k\oplus1}\rangle|b^2_{k\oplus1}\rangle)|c_{k\oplus s\oplus1}\rangle
\end{equation}
Charlie's photon path is now separated from the remaining photon paths. By doing so, Charlie is allowing the joint parties Bob$^1$ and Bob$^2$ to perform the joint operations. 
\item[Step4] At this stage, Bob$^1$ and Bob$^2$ jointly decide who will implement the operation first. Let Bob$^2$ will implement the operation first. Bob$^2$ is able to implement his operation $U^{2}_B$ if Bob$^1$ mixes the spatial states $|b^1_{k}\rangle$ and $|b^1_{k\oplus1}\rangle$ of his photon $\text{B}^1$ and lets one of the path (say $|b^1_{k}\rangle$) to interact with an arbitrary CS $|z\rangle$ via cross-kerr nonlinear interaction with interaction parameter $\theta$ and measures it. Consider the measurement outcome be $l=0$ $(1)$ corresponding to $|z\rangle$ $(|ze^{i\theta}\rangle)$. The new state after the measurement of Bob$^1$ is given as follows: 
\begin{equation}\label{eq:16}
    |\Xi_{kmnsl}\rangle=|a_{k\oplus m\oplus1}\rangle|b^1_{k\oplus l\oplus1}\rangle(\alpha|b^2_k\rangle+(-1)^{k\oplus m\oplus n\oplus s\oplus l}\beta|b^2_{k\oplus1}\rangle)|c_{k\oplus s\oplus1}\rangle
\end{equation}
It can be clearly seen from Eq. \ref{eq:16} that the coefficient $\alpha$ and $\beta$ which was initially with photon $X$ has shifted towards photon $\text{B}^2$. Bob$^2$ will now recover the state $\alpha|b^2_0\rangle+\beta|b^2_1\rangle$ by applying the appropriate unitary operation $Z^{k\oplus m\oplus n\oplus s\oplus l}_{S}X^{k}_{S}$, where $X_S=|b^{2}_{0}\rangle\langle b^{2}_{1}|+|b^{2}_{1}\rangle\langle b^{2}_{0}|$ and $Z_S=|b^{2}_{0}\rangle\langle b^{2}_{0}|-|b^{2}_{1}\rangle\langle b^{2}_{1}|$. Once Bob$^2$ recovers the state, then he will implement the operation $U^{2}_B$ on it, which will transform the state into the following: 
\begin{equation}\label{eq:17}
    |\Lambda_{kmsl}\rangle=|a_{k\oplus m\oplus1}\rangle|b^1_{k\oplus l\oplus1}\rangle(\alpha_{B^2}|b^2_0\rangle+\beta_{B^2}|b^2_1)|c_{k\oplus s\oplus1}\rangle
\end{equation}
\end{description}
Step3 and Step4 have been presented in pictorial form in Fig. \ref{fig:cjrio2}. 
\begin{description}
\item[Step5] The path of photon B$^1$ got separated in the previous step which we need to bring back into the spatial superposition by passing through a BBS. When $|b^{1}_{k\oplus l\oplus1}\rangle$ passes through a BBS, it triggers a new path $|b^{1}_{k\oplus l}\rangle$. Behind the BBS, Bob$^1$ picks one path (say $|b^{1}_{k\oplus l\oplus1}\rangle$) and lets it interact with an auxiliary CS $|z\rangle$ via cross-kerr interaction with interaction parameter $\theta$ and forwards the CS to Bob$^2$, which interacts with path $|b^{2}_{0}\rangle$ via cross-kerr interaction with interaction parameter $-\theta$. After the interaction, Bob$^2$ measures the CS, whose measurement outcome is $r=0$ $(1)$ corresponding to $|z\rangle$ $(|ze^{\pm i\theta})$. The measurement of Bob$^2$ here, transforms the state in Eq. \ref{eq:17} into a new state $|\Lambda_{kmslr}\rangle$ which can be written as follows:
\begin{equation}\label{eq:18}
     |\Lambda_{kmslr}\rangle=|a_{k\oplus m\oplus1}\rangle(\alpha_{B^2}|b^1_{k\oplus l\oplus r\oplus 1}\rangle|b^2_0\rangle+(-1)^{k\oplus l\oplus 1}\beta_{B^2}|b^1_{k\oplus l\oplus r}\rangle|b^2_1)|c_{k\oplus s\oplus1}\rangle
\end{equation}
\item[Step6] In the previous step, Bob$^1$ cooperates with Bob$^2$ to implement his operation $U^{2}_B$. Now its time for Bob$^2$ to cooperate with Bob$^1$ by mixing his photon path states $|b^{2}_{0}\rangle$ and $|b^{2}_{1}\rangle$ on a BBS and turning on the cross-kerr interaction between $|b^{2}_{1}\rangle$ and an auxiliary CS $|z\rangle$ with interaction parameter $\theta$. After the interaction, Bob$^2$ measures the CS whose measurement outcomes is $g=0$ $(1)$ corresponding to $|z\rangle$ $(|ze^{i\theta}\rangle)$. The measurement of Bob$^2$ transforms the state $|\Lambda_{kmslr}\rangle$ to $|\Lambda_{kmslrg}\rangle$ which is given as follows:
\begin{equation}
     |\Lambda_{kmslrg}\rangle=|a_{k\oplus m\oplus1}\rangle(\alpha_{B^2}|b^1_{k\oplus l\oplus r\oplus 1}\rangle+(-1)^{k\oplus l\oplus g\oplus 1}\beta_{B^2}|b^1_{k\oplus l\oplus r}\rangle)|b^2_g\rangle|c_{k\oplus s\oplus1}\rangle
\end{equation}
The coefficient $\alpha_{B^2}$ and $\beta_{B^2}$, which was initially with Bob$^2$ has now shifted to Bob$^1$. To recover the original path of photon $\text{B}^1$, Bob$^1$ will apply an appropriate unitary $Z^{k\oplus l\oplus g\oplus1}_{S}X^{k\oplus l\oplus r\oplus1}_{S}$ on his photon. After the recovery, the state will become
\begin{equation}
     |\Lambda^{'}_{kmslrg}\rangle=|a_{k\oplus m\oplus1}\rangle(\alpha_{B^2}|b^1_{0}\rangle+\beta_{B^2}|b^1_{1}\rangle)|b^2_g\rangle|c_{k\oplus s\oplus1}\rangle
\end{equation}
Bob$^1$ can now implement his unitary operation $U^1_B$ on his photon $\text{B}^1$, that transforms the state to $ |\Lambda^{''}_{kmslrg}\rangle$ which is given as
\begin{equation}
     |\Lambda^{''}_{kmslrg}\rangle=|a_{k\oplus m\oplus1}\rangle(\alpha_{B^1B^2}|b^1_0\rangle+\beta_{B^1B^2}|b^1_1\rangle)|b^2_g\rangle|c_{k\oplus s\oplus1}\rangle
\end{equation}
It is noted that $\alpha_{B^1B^2}|b^1_0\rangle+\beta_{B^1B^2}|b^1_1\rangle=U^{1}_{B}U^{2}_{B}(\alpha|b^{1}_{0}+\beta|b^{1}_{1})$. But the task is not completed yet, the coefficients $\alpha_{B^1B^2}$ and $\beta_{B^1B^2}$ need to be transferred to Alice' node. To do so, the communicating parties will now use their P-DOF which is described in Sec. \ref{P-DOF}.
\end{description}
Step5 and Step6 have been presented in pictorial form in Fig. \ref{fig:cjrio3}. 
\begin{figure}
    \centering
    \includegraphics[width=\textwidth]{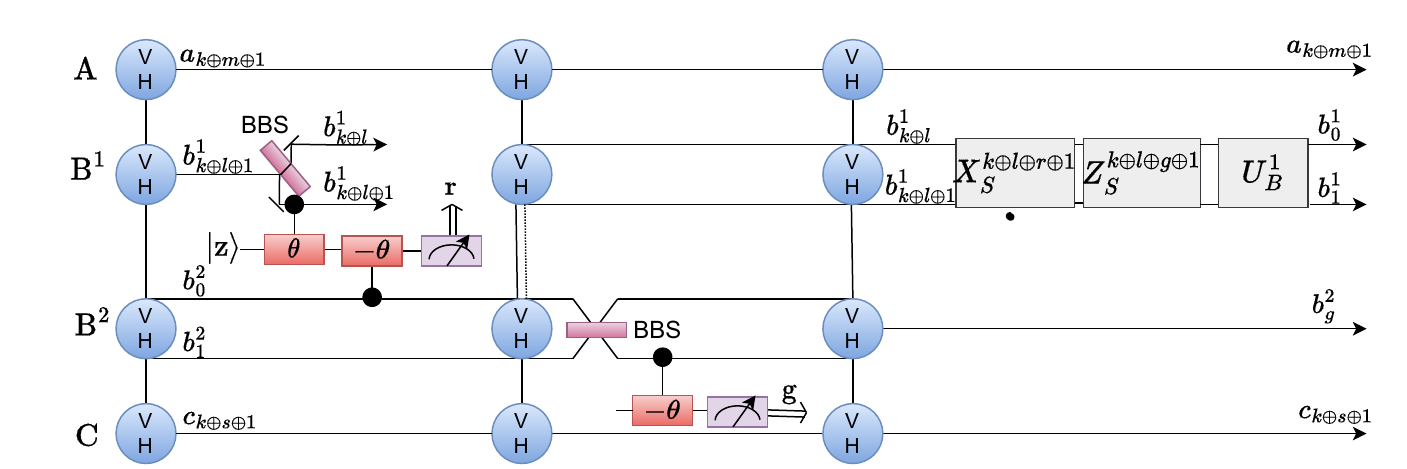}
    \caption{(Color online) A schematic which represents Step5 and Step6 of the CJRIO protocol. Here, Bob$^1$ first triggers a new path using a BBS and allows the non-linear interaction $K_{b^1_{k\oplus l\oplus1}}(\theta)|z\rangle|b^1_{k\oplus l\oplus1}\rangle$ and forwards it to Bob$^2$ which allows the interaction $K_{b^2_0}(-\theta)|z\rangle|b^2_0\rangle$ and measures the CS, whose measurement outcome is $r$. Bob$^2$ then mixes two spatial paths of his photon and turns on proper cross-kerr interaction between a CS and one path of his photon followed by measuring the CS with outcome $g$ as shown in the figure, that allows Bob$^1$ to implement an appropriate unitaries to get $\alpha_{B^1B^2}|b^1_0\rangle+\beta_{B^1B^2}|b^1_1\rangle$.}
    \label{fig:cjrio3}
\end{figure}
\subsection{Utilizing P-DOF\label{P-DOF}} 

Taking into account the P-DOF, the combined state after Step6 can now be written as following:
\begin{equation}\label{eq:22}
    |\phi^{P}\rangle=|\Lambda^{''}_{kmslrg}\rangle|Q^{P}\rangle
\end{equation}
The expanded form of Eq. \ref{eq:22} can be written as
\begin{equation}
\begin{split}
     |\phi^{P}\rangle &=|a_{k\oplus m\oplus1}\rangle[\alpha_{B^1B^2}|\text{H}\rangle_{A}|\text{H},b^1_0\rangle_{B^1}|\text{H},b^2_g\rangle_{B^2}|\text{H}\rangle_{C}+\alpha_{B^1B^2}|\text{V}\rangle_{A}|\text{V},b^1_0\rangle_{B^1}|\text{V},b^2_g\rangle_{B^2}|\text{V}\rangle_C\\
     &+\beta_{B^1B^2}|\text{H}\rangle_{A}|\text{H},b^1_1\rangle_{B^1}|\text{H},b^2_g\rangle_{B^2}|\text{H}\rangle_{C}+\beta_{B^1B^2}|\text{V}\rangle_{A}|\text{V},b^1_1\rangle_{B^1}|\text{V},b^2_g\rangle_{B^2}|\text{V}\rangle_C]|c_{k\oplus s\oplus1}\rangle
\end{split}
\end{equation}
Here, $|H,b^1_0\rangle$ denotes state of horizontally polarized photon propagating along path $b^1_0$ and similarly for $|H,b^2_g\rangle$, $|V,b^1_0\rangle$, $|V,b^2_g\rangle$, $|H,b^1_1\rangle$, $|H,b^2_g\rangle$, $|V,b^1_1\rangle$ and $|V,b^2_g\rangle$.\\

\begin{figure}
    \centering
    \includegraphics[width=\textwidth]{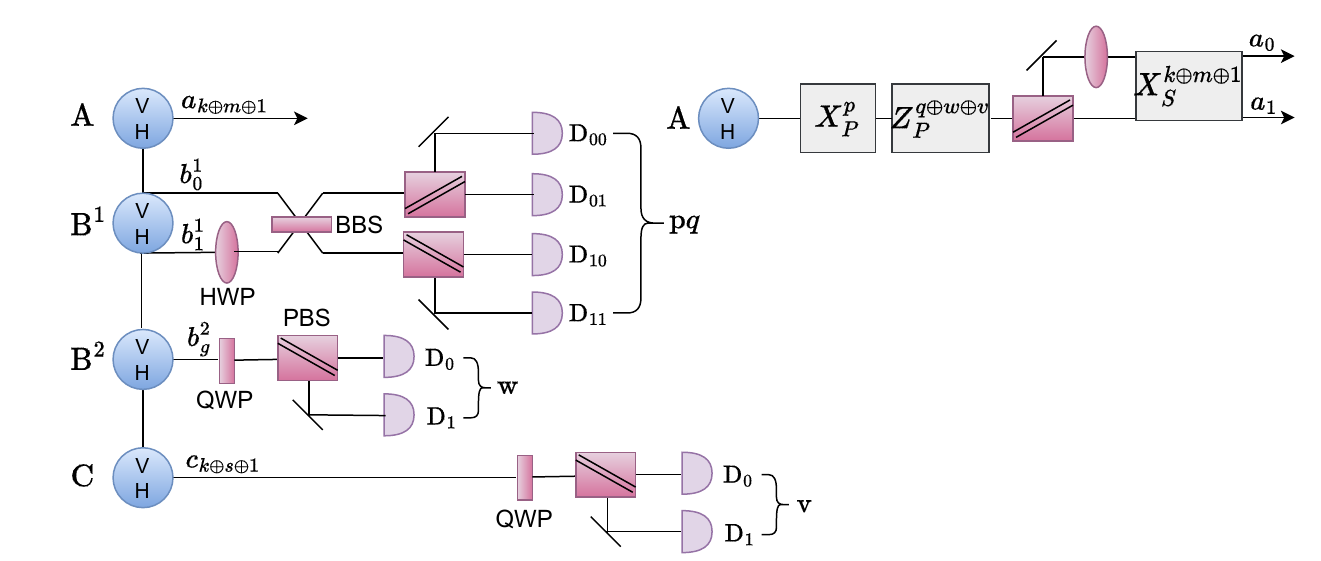}
    \caption{(Color Online) A schematic which represents Step7 to Step9 of the CJRIO protocol. Here, the joint parties Bob$^1$ and Bob$^2$ and the controller Charlie measure their photons in an appropriate basis using HWP, QWP, BBS and PBS. After the measurement photons $\text{B}^1$, $\text{B}^2$ and $\text{C}$ collapsed and we are only left with photon $\text{A}$. Alice then applies appropriate unitary followed by PBS and HWP in one path to obtain the desired state $\alpha_{B^1B^2}|a_{0}\rangle+\beta_{B^1B^2}|a_{1}\rangle$.}
    \label{fig:cjrio4}
\end{figure}

The initial goal was to implement an operator jointly at Alice's node. To achieve the goal, the coefficients $\alpha_{B^1B^2}$ and $\beta_{B^1B^2}$ have to be shifted towards Alice. For that, the joint parties (Bob$^1$, Bob$^2$) and the controller (Charlie) will have to measure their photons in appropriate bases.
\begin{description}
\item[Step7] First, the joint parties Bob$^1$ and Bob$^2$ measure their photons in an appropriate bases. Before the measurement, Bob$^1$ puts a half-wave plate (HWF) on path $b^1_1$ to exchange the photon polarization $|H,b^1_1\rangle\rightleftarrows|V,b^1_1\rangle$ and then mixes the two paths $|b^1_0\rangle$ and $|b^1_1\rangle$ of his photon on a BBS, which transforms the state to the following:
\begin{equation}
    \begin{split}
       |\Omega\rangle& =|a_{k\oplus m\oplus1}\rangle[\alpha_{B^1B^2}|\text{H}\rangle_{A}|\text{H},b^1_0\rangle_{B^1}|\text{H},b^2_g\rangle_{B^2}|\text{H}\rangle_{C}+\alpha_{B^1B^2}|\text{H}\rangle_{A}|\text{H},b^1_1\rangle_{B^1}|\text{H},b^2_g\rangle_{B^2}|\text{H}\rangle_{C}\\
       &+\alpha_{B^1B^2}|\text{V}\rangle_{A}|\text{V},b^1_0\rangle_{B^1}|\text{V},b^2_g\rangle_{B^2}|\text{V}\rangle_C+\alpha_{B^1B^2}|\text{V}\rangle_{A}|\text{V},b^1_1\rangle_{B^1}|\text{V},b^2_g\rangle_{B^2}|\text{V}\rangle_C\\
       &+\beta_{B^1B^2}|\text{H}\rangle_{A}|\text{V},b^1_0\rangle_{B^1}|\text{H},b^2_g\rangle_{B^2}|\text{H}\rangle_{C}-\beta_{B^1B^2}|\text{H}\rangle_{A}|\text{V},b^1_1\rangle_{B^1}|\text{H},b^2_g\rangle_{B^2}|\text{H}\rangle_{C}\\
       &+\beta_{B^1B^2}|\text{V}\rangle_{A}|\text{H},b^1_0\rangle_{B^1}|\text{V},b^2_g\rangle_{B^2}|\text{V}\rangle_C-\beta_{B^1B^2}|\text{V}\rangle_{A}|\text{H},b^1_1\rangle_{B^1}|\text{V},b^2_g\rangle_{B^2}|\text{V}\rangle_C]|c_{k\oplus s\oplus1}\rangle
    \end{split}
\end{equation}
Bob$^2$ now puts a quarter wave plate (QWP) on path $|b^2_g\rangle$ to, up to the normalization factor, transform $|H,b^2_g\rangle$ to $(|H,b^2_g\rangle+|V,b^2_g\rangle)$ and $|V,b^2_g\rangle$ to $|H,b^2_g\rangle-|V,b^2_g\rangle$. The transformed state is given as follows:
\begin{equation}
    \begin{split}
        |\Omega^{'}\rangle&=|a_{k\oplus m\oplus1}\rangle[|\text{H},b^1_0\rangle_{B^1}|\text{H},b^2_g\rangle_{B^2}(\alpha_{B^1B^2}|\text{H}\rangle_{A}|\text{H}\rangle_{C}+\beta_{B^1B^2}|\text{V}\rangle_{A}|\text{V}\rangle_{C})\\
        &+|\text{H},b^1_0\rangle_{B^1}|\text{V},b^2_g\rangle_{B^2}(\alpha_{B^1B^2}|\text{H}\rangle_{A}|\text{H}\rangle_{C}-  \beta_{B^1B^2}\text{V}\rangle_{A}|\text{V}\rangle_{C})\\
        &+|\text{H},b^1_1\rangle_{B^1}|\text{H},b^2_g\rangle_{B^2}(\alpha_{B^1B^2}|\text{H}\rangle_{A}|\text{H}\rangle_{C}-\beta_{B^1B^2}|\text{V}\rangle_{A}|\text{V}\rangle_{C})\\
        &+|\text{H},b^1_1\rangle_{B^1}|\text{V},b^2_g\rangle_{B^2}(\alpha_{B^1B^2}|\text{H}\rangle_{A}|\text{H}\rangle_{C}+\beta_{B^1B^2}|\text{V}\rangle_{A}|\text{V}\rangle_{C})\\
        &+|\text{V},b^1_0\rangle_{B^1}|\text{H},b^2_g\rangle_{B^2}(\alpha_{B^1B^2}|\text{V}\rangle_{A}|\text{V}\rangle_{C}+\beta_{B^1B^2}|\text{H}\rangle_{A}|\text{H}\rangle_{C})\\
        &-|\text{V},b^1_0\rangle_{B^1}|\text{V},b^2_g\rangle_{B^2}(\alpha_{B^1B^2}|\text{V}\rangle_{A}|\text{V}\rangle_{C}-\beta_{B^1B^2}|\text{H}\rangle_{A}|\text{H}\rangle_{C})\\
        &+|\text{V},b^1_1\rangle_{B^1}|\text{H},b^2_g\rangle_{B^2}(\alpha_{B^1B^2}|\text{V}\rangle_{A}|\text{V}\rangle_{C}-\beta_{B^1B^2}|\text{H}\rangle_{A}|\text{H}\rangle_{C})\\
        &-|\text{V},b^1_1\rangle_{B^1}|\text{V},b^2_g\rangle_{B^2}(\alpha_{B^1B^2}|\text{V}\rangle_{A}|\text{V}\rangle_{C}+\beta_{B^1B^2}|\text{H}\rangle_{A}|\text{H}\rangle_{C})]|c_{k\oplus s\oplus1}\rangle
    \end{split}
\end{equation}
Now Bob$^1$ and Bob$^2$ will measure their photons in an appropriate basis. Bob$^1$ (Bob$^2$) measures his photon $\text{B}^1$ $(\text{B}^2)$ in the basis $\{|\text{H},b^1_0\rangle,|\text{H},b^1_1\rangle,|\text{V},b^1_0\rangle,|\text{V},b^1_1\rangle\}$ $(\{|\text{H},b^2_g\rangle,|\text{V},b^2_g\rangle\})$, whose corresponding measurement results are $pq=00$, $01$, $10$, $11$ ($w=0$, $1$). The collapsed state after the measurement is given as follows:
\begin{equation}
     |\Omega_{kmpqw}\rangle=|a_{k\oplus m\oplus1}\rangle\left\{\begin{matrix}
(\alpha_{B^1B^2}|\text{H}\rangle_{A}|\text{H}\rangle_{C}+\beta_{B^1B^2}|\text{V}\rangle_{A}|\text{V}\rangle_{C})|c_{k\oplus s\oplus1}\rangle & \text{for} & pqw= 000,011\\ 
(\alpha_{B^1B^2}|\text{H}\rangle_{A}|\text{H}\rangle_{C}-\beta_{B^1B^2}|\text{V}\rangle_{A}|\text{V}\rangle_{C})|c_{k\oplus s\oplus1}\rangle & \text{for} & pqw= 010,001\\ 
(\alpha_{B^1B^2}|\text{V}\rangle_{A}|\text{V}\rangle_{C}+\beta_{B^1B^2}|\text{H}\rangle_{A}|\text{H}\rangle_{C})|c_{k\oplus s\oplus1}\rangle & \text{for} & pqw= 100,111\\ 
(\alpha_{B^1B^2}|\text{V}\rangle_{A}|\text{V}\rangle_{C}-\beta_{B^1B^2}|\text{H}\rangle_{A}|\text{H}\rangle_{C})|c_{k\oplus s\oplus1}\rangle & \text{for} & pqw= 110,101
\end{matrix}\right.
\end{equation}
\item[Step8] The controller Charlie once again uses her power. If she wants to stop the protocol, she does nothing otherwise she places a QWP on her photon path $c_{k\oplus s\oplus1}$ to rotate the polarization state of her photon state from $|H\rangle_C\rightarrow|H\rangle_C+|V\rangle_C$ and $|V\rangle_C\rightarrow|H\rangle_C-|V\rangle_C$ (normalization is omitted). After that, Charlie lets her photon pass through a polarizing beam splitter (PBS), which transmits photon of horizontal polarization but reflects that of vertical one, and measures it in the basis $\{|\text{H},c_{k\oplus s\oplus1}\rangle, |\text{V},c_{k\oplus s\oplus1}\rangle$, whose corresponding measurement outcome is $v=0$, $1$. The collapsed state after Charlie's measurement is given as follows:
\begin{equation}
   |\Omega_{kmpqwv}\rangle=|a_{k\oplus m\oplus1}\rangle\left\{\begin{matrix}
(\alpha_{B^1B^2}|\text{H}\rangle+\beta_{B^1B^2}|\text{V}\rangle) & \text{for} & pqwv= 0000,0101, 0011, 0110\\ 
(\alpha_{B^1B^2}|\text{H}\rangle-\beta_{B^1B^2}|\text{V}\rangle) & \text{for} & pqwv= 0001, 0100, 0010, 0111\\ 
(\alpha_{B^1B^2}|\text{V}\rangle+\beta_{B^1B^2}|\text{H}\rangle) & \text{for} & pqwv= 1000, 1101, 1011, 1110\\ 
(\alpha_{B^1B^2}|\text{V}\rangle-\beta_{B^1B^2}|\text{H}\rangle) & \text{for} & pqwv= 1001, 1100, 1010, 1111
\end{matrix}\right.
\end{equation}
It can be seen that the coefficients $\alpha_{B^1B^2}$ and $\beta_{B^1B^2}$ are finally shifted to Alice. Now, Alice applies $Z^{q\oplus w\oplus v}_{P}X^{p}_{P}$ on her photon, where $X_P=|H\rangle\langle V|+|V\rangle\langle H|$ and $Z_P=|H\rangle\langle H|-|V\rangle\langle V|$, to obtain a new state given as follow:
\begin{equation}\label{eq:28}
    |\Omega_{km}\rangle=(\alpha_{B^1B^2}|H\rangle+\beta_{B^1B^2}|V\rangle)|a_{k\oplus m\oplus1}\rangle
\end{equation}
Now, the next step is to transform Alice's photon state from P-DOF to the photon's state in S-DOF.
\item[Step9] In this step, Alice first applies PBS on the state $|\Omega_{km}\rangle$, which triggers a new path and turns the state into $(\alpha_{B^1B^2}|H,a_{k\oplus m\oplus1}\rangle+\beta_{B^1B^2}|V,a_{k\oplus m}\rangle)$. Further, a HWP is placed in one of the path (say $a_{k\oplus m\oplus1}$) which generates a new state $(\alpha_{B^1B^2}|a_{k\oplus m\oplus1}\rangle+\beta_{B^1B^2}|a_{k\oplus m}\rangle)|V\rangle$. Alice finally applies an operator $X^{k\oplus m\oplus1}_{S}$, which recovers the required state $(\alpha_{B^1B^2}|a_{0}\rangle+\beta_{B^1B^2}|a_{1}\rangle)=U_{B}^1U_{B}^2|\psi\rangle_A$. The task of CJRIO has been successfully achieved now. 
\end{description}
The Step7, Step8 and Step9 have been presented in pictorial form in Fig. \ref{fig:cjrio4}. 
\section{A Possible Generalization for CJRIO}\label{sec:generalization}
The proposed scheme for CJRIO can be generalized to $M$-joint parties (say Bob$^1$, Bob$^2$,..., Bob$^M$) and $N$-controllers (say Charlie$^1$, Charlie$^2$,...,Charlie$^N$). The operator of the respective (say $i^{th})$ joint party is $U^{i}_B$ which is given as follows:
\begin{equation}
    U^i_B=\begin{pmatrix}
    u^i_B & v^i_B\\ 
    -v^{*i}_B & u^{*i}_B
    \end{pmatrix}
\end{equation}
The task here is to jointly prepare an arbitrary operation on Alice's photon by all $M$ parties which is mathematically represented as
\begin{equation}
    \begin{split}
    |\psi_{B^1B^2...B^M}\rangle&=U^{1}_BU^{2}_B...U^{M}_B|\psi\rangle_X\\
    &=(\alpha_{B^1B^2...B^M}|x_0\rangle+\beta_{B^1B^2...B^M}|x_1\rangle)
    \end{split}
\end{equation}
The quantum channel used to complete the task of CJRIO for $M$-joint parties and $N$-controllers can be given as: 
\begin{equation}
    |Q\rangle_{AB^1B^2...B^MC^1C^2...C^N}=|Q^{S}\rangle_{AB^1B^2...B^MC^1C^2...C^N}|Q^{P}\rangle_{AB^1B^2...B^MC^1C^2...C^N}
\end{equation}
where (up to the normalization factor)
\begin{equation}
    |Q^{S}\rangle_{AB^1B^2...B^MC^1C^2...C^N}=|a_0\rangle_A|b^{1}_0\rangle_{B^1}|b^{2}_0\rangle_{B^2}...|b^{M}_0\rangle_{B^M}|c^{1}_0\rangle_{C^1}|c^{2}_0\rangle_{C^2}...|c^{N}_0\rangle_{C^N}+|a_1\rangle_{A}|b^{1}_1\rangle_{B^1}|b^{2}_1\rangle_{B^2}...|b^{M}_1\rangle_{B^M}|c^{1}_1\rangle_{C^1}|c^{2}_1\rangle_{C^2}...|c^{N}_1\rangle_{C^N}
\end{equation}
\begin{equation}
    |Q^{P}\rangle_{AB^1B^2...B^MC^1C^2...C^N}=|H\rangle_A|H\rangle_{B^1}|H\rangle_{B^2}...|H\rangle_{B^M}|H\rangle_{C^1}|H\rangle_{C^2}...|H\rangle_{C^N}+|V\rangle_A|V\rangle_{B^1}|V\rangle_{B^2}...|V\rangle_{B^M}|V\rangle_{C^1}|V\rangle_{C^2}...|V\rangle_{C^N}
\end{equation}
The steps involved to achieve the task of CJRIO for $M$-joint parties and $N$-controllers, are described as follows. 
\begin{description}
    \item[Step1] This step is same as Step1 of Sec. \ref{S-DOF} where Alice's unknown state $|\psi\rangle_X$ gets entangled with the quantum channel $|Q^{S}\rangle_{AB^1B^2...B^MC^1C^2...C^N}$ in S-DOF. The entangled state can be shown as follows:
    \begin{equation}
        |\Phi_k\rangle=\alpha|x_0\rangle|a_k\rangle\bigotimes_{i=1}^{M}|b^{i}_k\rangle\bigotimes_{j=1}^{N}|c^{j}_{k}\rangle+\beta|x_1\rangle|a_{k\oplus1}\rangle\bigotimes_{i=1}^{M}|b^{i}_{k\oplus1}\rangle\bigotimes_{j=1}^{N}|c^{j}_{k\oplus1}\rangle
    \end{equation}
    \item[Step2] In this step, Alice tries to disentangle her photons from the remaining photons in the quantum channel in a similar manner as Step2 of Sec. \ref{S-DOF}. The collapsed state is given as follows:
    \begin{equation}
        |\Phi_{kmn}\rangle=|a_{k\oplus m\oplus1}\rangle(\alpha\bigotimes_{i=1}^{M}|b^{i}_k\rangle\bigotimes_{j=1}^{N}|c^{j}_{k}\rangle+\beta\bigotimes_{i=1}^{M}|b^{i}_{k\oplus1}\rangle\bigotimes_{j=1}^{N}|c^{j}_{k\oplus1}\rangle)
    \end{equation}
    \item[Step3] Each controller mixes her photon paths on a BBS and then lets one path of photon interact with a CS via cross-kerr interaction with interaction parameter $\theta$ and measures it, whose measurement outcomes are $s_j=0$ $(1)$ corresponding to $|z\rangle$ $(|ze^{i\theta}\rangle)$. After the measurement of each controller, the state turns into the following:
    \begin{equation}
        |\Phi_{kmns}\rangle=|a_{k\oplus m\oplus1}\rangle(\alpha\bigotimes^{M}_{i=1}|b^{i}_{k}\rangle-(-1)^{m\oplus n\oplus s_{1}\oplus s_{2}\oplus...\oplus s_{N}}\beta\bigotimes^{M}_{i=1}|b^{i}_{k\oplus1}\rangle)\bigotimes^{N}_{j=1}|c_{k\oplus s_{j}\oplus1}\rangle
    \end{equation}
    \item[Step4] In the previous step, controllers allow joint parties to complete the task. Here comes the role of joint parties to implement their respective operation $U^{i}_{B}$, where $(i=1,2,..., M)$. The joint parties decide among themselves who will implement the operation first. Let them decide that Bob$^M$ will implement his operation first then each of the remaining joint parties follows the Step4 of Sec. \ref{S-DOF}, which gives the measurement outcomes $l_i=0$ $(1)$. Bob$^M$ then applies an appropriate unitary to get $\alpha|b^{M}_0\rangle+\beta|b^{M}_1\rangle$. Bob$^M$ is now ready to implement his operator $U^{M}_{B}$ to get $\alpha_{B^M}|b^{M}_0\rangle+\beta_{B^M}|b^{M}_1\rangle$.
    \item[Step5] In this step, the coefficient $\alpha_{B^M}$ and $\beta_{B^M}$, which are with Bob$^{M}$, are gradually shifted towards Bob$^{M-1}$, Bob$^{M-2}$,...and so on. The two alternate joint parties work together at once here. First, suppose Bob$^{M}$ and Bob$^{M-1}$ work together. Bob$^{M-1}$ places a BBS on his photon path which triggers a new path and then allows one of the paths to interact with an auxiliary CS $|z\rangle$ via crosss-kerr nonlinear interaction of parameter $\theta$ and forwards it to Bob$^{M}$, which he allows to interact with his one of photon path and measures it, whose measurement outcome is $r_M=0$ $(1)$. Bob$^{M}$ now mixes two paths of his photon on a BBS and allows one path to interact with an auxiliary CS via cross-kerr nonlinear interaction with interaction parameter $\theta$ and measures it, whose measurement outcome is $g_{M}=0$ $(1)$. Depending on the measurement outcomes, Bob$^{M-1}$ will apply an appropriate unitary to get $\alpha_{B^M}|b^{M-1}_0\rangle+\beta_{B^M}|b^{M-1}_1\rangle$, on which he will operate $U^{M-1}_{B}$ to get $\alpha_{B^{M-1}B^M}|b^{M-1}_0\rangle+\beta_{B^{M-1}B^M}|b^{M-1}_1\rangle$. This process is repeated now for Bob$^{M-1}$ and Bob$^{M-2}$, and so on till Bob$^{1}$ and Bob$^{2}$. The final state will now become
    \begin{equation}
        |\Phi_{kmns_{j}l_{i}r_{i}g_{i}}\rangle=|a_{k\oplus m\oplus1}\rangle(\alpha_{B^1B^2...B^M}|b^{1}_{0}\rangle+\beta_{B^1B^2...B^M}|b^{1}_{1}\rangle)\bigotimes^{M}_{i=2}|b^{i}_{g_{i}}\rangle\bigotimes^{N}_{j=1}|c_{k\oplus s_{j}\oplus1}\rangle
    \end{equation}
\item[Step6] The joint parties take part in this step. They measure their photons in an appropriate bases and hand over the task to the sender and controllers. As we know from the previous step only Bob$^{1}$'s photon is in spatial superposition and the rest joint parties' photons are spatially separated. So, Bob$^{1}$ places a HWF on path $b^{1}_0$ and mixes the superimposed path on a BBS, rest of the joint parties put a QWP on their photons path. All the joint parties now measure their photons in an appropriate basis. The role of joint parties ended here.  
\item[Step7] All the controllers now put a QWP in their photons path and pass it through PBS and measure it in an appropriate bases. The controllers roles are ended here. Now, Alice will apply an appropriate Pauli operations to get $(\alpha_{B^1B^2...B^M}|\text{H}\rangle+\beta_{B^1B^2...B^M}|\text{V}\rangle)|a_{k\oplus m\oplus1}\rangle$, which is same as Eq. \ref{eq:28}. 
\item[Step8] This step is same as Step9 of Sec. \ref{P-DOF}. 
\end{description}
\section{Existing variants of RIO as a special case of CJRIO\label{RIOasCJRIO}}
The proposed scheme for CJRIO can be seen as a generalized scheme and all existing variants of RIO scheme can be obtained as a special case of the proposed scheme. e.g. if one removes Charlie and the corresponding steps from our proposed scheme, then our scheme reduces to the existing scheme for JRIO reported in reference \cite{BAn22}. Removing Charlie will first reduce the quantum channel in Eq. \ref{eq:QC} to three qubit hyperentangled state and then remove the corresponding steps that involve Charlie, which are Step3 and Step8 of Sec. \ref{S-DOF}. Similarly, if one removes either of the joint parties Bob$^1$ or Bob$^2$ from our proposed scheme, then our scheme reduces to the existing scheme for CRIO reported in reference \cite{BB22}. Let's remove Bob$^2$, this changes the quantum channel in Eq. \ref{eq:QC} and the corresponding Steps which are Step4, Step5, Step6 and Step7. Step4 is removed which retains the spatial superposition of photon $\text{B}^1$ from the previous step, on which Bob$^1$ applies an appropriate unitary and his operation to get $\alpha_{B^1}|b^1_0\rangle+\beta_{B^1}|b^1_1\rangle$. Now there is no need of Step5 and Step6 and the role of Bob$^2$ is removed from Step7, which ended up with a scheme for CRIO.
\section{Discussion and Conclusion}\label{sec:conclusion}
In this work, we provide 2 interesting schemes for CJRIO which can be viewed as the basic building blocks for distributed photonic quantum computing and deserves particular use in the noisy intermediate-scale quantum era when scalable quantum computers have not yet been available. Specifically, here we first propose a scheme for CJRIO that allows two users to jointly prepare an arbitrary unitary operation on an unknown state at remote node in presence of a controller. The proposed scheme is completed using a four-qubit hyperentangled state, which is entangled in both S-DOF and P-DOF of photons. Finally, the idea is generalized to propose a scheme that allows an arbitrary number of joint parties as well as of controllers to perform the CJRIO task. As all these schemes are designed considering their realization using photonic quantum states and as it's described in the introduction that the distributed computing requires CJRIO in the implementations involving master-slave architecture, the present work seems to be very useful in distributed photonic quantum computing. Further, before we conclude this paper, it may be apt to note that the proposed schemes are the first set of schemes for CJRIO and seem to be experimentally realizable with the existing technology. Keeping the above in mind, we conclude this paper with a hope that the work will be experimentally realized and found applications in distributed photonic quantum computing in the near future.
\section*{Acknowledgment}
Authors acknowledge the support from the QUEST scheme of Interdisciplinary
Cyber Physical Systems (ICPS) program of the Department of Science
and Technology (DST), India (Grant No.: DST/ICPS/QuST/Theme-1/2019/14
(Q80)). 

\section*{Data Availability} Data sharing not applicable to this article as no datasets were generated or analysed during the current study.

\bibliographystyle{unsrt}
\bibliography{cjrio}

\begin{thebibliography}{10}

\bibitem{BB93}
Charles~H Bennett, Gilles Brassard, Claude Cr{\'e}peau, Richard Jozsa, Asher Peres, and William~K Wootters.
\newblock Teleporting an unknown quantum state via dual classical and einstein-podolsky-rosen channels.
\newblock {\em Physical Review Letters}, 70(13):1895, 1993.

\bibitem{P00}
Arun~K Pati.
\newblock Minimum classical bit for remote preparation and measurement of a qubit.
\newblock {\em Physical Review A}, 63(1):014302, 2000.

\bibitem{SSB+15}
Vishal Sharma, Chitra Shukla, Subhashish Banerjee, and Anirban Pathak.
\newblock Controlled bidirectional remote state preparation in noisy environment: a generalized view.
\newblock {\em Quantum Information Processing}, 14:3441--3464, 2015.

\bibitem{HVC+01}
Susana~F Huelga, Joan~A Vaccaro, Anthony Chefles, and Martin~B Plenio.
\newblock Quantum remote control: teleportation of unitary operations.
\newblock {\em Physical Review A}, 63(4):042303, 2001.

\bibitem{TVP15}
Kishore Thapliyal, Amit Verma, and Anirban Pathak.
\newblock A general method for selecting quantum channel for bidirectional controlled state teleportation and other schemes of controlled quantum communication.
\newblock {\em Quantum Information Processing}, 14:4601--4614, 2015.

\bibitem{SST+17}
Mitali Sisodia, Abhishek Shukla, Kishore Thapliyal, and Anirban Pathak.
\newblock Design and experimental realization of an optimal scheme for teleportation of an n-qubit quantum state.
\newblock {\em Quantum Information Processing}, 16:292, 2017.

\bibitem{HPV02}
Susana~F Huelga, Martin~B Plenio, and Joan~A Vaccaro.
\newblock Remote control of restricted sets of operations: teleportation of angles.
\newblock {\em Physical Review A}, 65(4):042316, 2002.

\bibitem{A07}
Nguyen~Ba An.
\newblock Remote application of hidden operators.
\newblock {\em Physics Letters A}, 364(3-4):198--202, 2007.

\bibitem{FL08}
QiuBo Fan and DongDong Liu.
\newblock Controlled remote implementation of partially unknown quantum operation.
\newblock {\em Science in China Series G: Physics, Mechanics and Astronomy}, 51:1661--1667, 2008.

\bibitem{PH19}
Jia-Yin Peng and Yong He.
\newblock Cyclic controlled remote implementation of partially unknown quantum operations.
\newblock {\em International Journal of Theoretical Physics}, 58:3065--3072, 2019.

\bibitem{PY+22}
Jia-yin Peng, Zhen Yang, Liang Tang, and Jia-sheng Peng.
\newblock Double-direction cyclic controlled remote implementation of partially known quantum operations.
\newblock {\em International Journal of Theoretical Physics}, 61(10):256, 2022.

\bibitem{XLG05}
Guo-Yong Xiang, Jian Li, and Guang-Can Guo.
\newblock Teleporting a rotation on remote photons.
\newblock {\em Physical Review A}, 71(4):044304, 2005.

\bibitem{BB22}
Nguyen~Ba An and Bich~Thi Cao.
\newblock Controlled remote implementation of operators via hyperentanglement.
\newblock {\em Journal of Physics A: Mathematical and Theoretical}, 55(22):225307, 2022.

\bibitem{BAn22}
Nguyen~Ba An.
\newblock Joint remote implementation of operators.
\newblock {\em Journal of Physics A: Mathematical and Theoretical}, 55(39):395304, 2022.

\bibitem{WG24}
Meiyu Wang and Hao Guo.
\newblock Quantum remote control utilizing multiple degrees of freedom.
\newblock {\em Optics \& Laser Technology}, 169:110075, 2024.

\bibitem{QC23}
Xinyu Qiu and Lin Chen.
\newblock Controlled remote implementation of operations via graph states.
\newblock {\em Annalen der Physik}, 535(12):2300320, 2023.

\bibitem{GP07}
Manu Gupta and Anirban Pathak.
\newblock A scheme for distributed quantum search through simultaneous state transfer mechanism.
\newblock {\em Annalen der Physik}, 519(12):791--797, 2007.

\bibitem{CCC20}
Daniele Cuomo, Marcello Caleffi, and Angela~Sara Cacciapuoti.
\newblock Towards a distributed quantum computing ecosystem.
\newblock {\em IET Quantum Communication}, 1(1):3--8, 2020.

\bibitem{MFP+19}
Marco Minelli, Federica Ferraguti, Nicola Piccinelli, Riccardo Muradore, and Cristian Secchi.
\newblock An energy-shared two-layer approach for multi-master-multi-slave bilateral teleoperation systems.
\newblock In {\em 2019 International Conference on Robotics and Automation (ICRA)}, pages 423--429. IEEE, 2019.

\bibitem{DR+17Hyperentangle}
Fu-Guo Deng, Bao-Cang Ren, and Xi-Han Li.
\newblock Quantum hyperentanglement and its applications in quantum information processing.
\newblock {\em Science bulletin}, 62(1):46--68, 2017.

\bibitem{YC24Hyperentangle}
Yiqian Yang and Liangcai Cao.
\newblock Quantum hyper-entangled system with multiple qubits based on spontaneous parametric down-conversion and birefringence effect.
\newblock {\em Optical and Quantum Electronics}, 56(1):12, 2024.

\bibitem{NL+23Hyperentangle}
Liat Nemirovsky-Levy, Mark Lubarov, and Mordechai Segev.
\newblock Generation of two-dimensional cluster states using hyperentanglement.
\newblock In {\em Quantum 2.0}, pages QTu4A--3. Optica Publishing Group, 2023.

\bibitem{ZY+23Hyperentangle}
Peng Zhao, Meng-Ying Yang, Sha Zhu, Lan Zhou, Wei Zhong, Ming-Ming Du, and Yu-Bo Sheng.
\newblock Generation of hyperentangled state encoded in three degrees of freedom.
\newblock {\em Science China Physics, Mechanics \& Astronomy}, 66(10):100311, 2023.

\end{thebibliography}

\end{document}